\documentclass[prb,twocolumn,showpacs,preprintnumbers,amsmath,amssymb]{revtex4}
\usepackage{graphicx}% Include figure files
\usepackage{dcolumn}% Align table columns on decimal point
\begin{document}

% Use the \preprint command to place your local institutional report
% number in the upper righthand corner of the title page in preprint mode.
% Multiple \preprint commands are allowed.
% Use the 'preprintnumbers' class option to override journal defaults
% to display numbers if necessary
%\preprint{}

%Title of paper
\title{ Anomalous low temperature state of CeOs$_4$Sb$_{12}$: Magnetic field and La-impurity study }

% repeat the \author .. \affiliation  etc. as needed
% \email, \thanks, \homepage, \altaffiliation all apply to the current
% author. Explanatory text should go in the []'s, actual e-mail
% address or url should go in the {}'s for \email and \homepage.
% Please use the appropriate macro foreach each type of information

% \affiliation command applies to all authors since the last
% \affiliation command. The \affiliation command should follow the
% other information
% \affiliation can be followed by \email, \homepage, \thanks as well.
\author{C. R. Rotundu and B. Andraka}
\email[]{andraka@phys.ufl.edu}
%\homepage[]{Your web page}
%\thanks{}
%\altaffiliation{}
\affiliation{ Department of Physics, University of Florida\\
P.O. Box 118440, Gainesville, Florida  32611-8440, USA }

%Collaboration name if desired (requires use of superscriptaddress
%option in \documentclass). \noaffiliation is required (may also be
%used with the \author command).
%\collaboration can be followed by \email, \homepage, \thanks as well.
%\collaboration{}
%\noaffiliation

\date{\today}

\begin{abstract}
Specific heat for single crystalline samples of Ce$_{1-x}$La$_x$Os$_4$Sb$_{12}$ at zero-field and magnetic fields to 14 T is reported. Our results confirm enhanced value of the electronic specific heat coefficient in the paramagnetic state. They provide arguments for the intrinsic origin of the 1.1 K anomaly. This transition leads to opening of the gap at the Fermi surface. This low temperature state of CeOs$_4$Sb$_{12}$ is extremely sensitive to chemical impurities. 2 \% of La substituted for Ce suppresses the transition and reduces the electronic specific heat coefficient. The magnetic field response of the specific heat is also anomalous. 
\end{abstract}

% insert suggested PACS numbers in braces on next line
% insert suggested keywords - APS authors don't need to do this
%\keywords{}

%\maketitle must follow title, authors, abstract, \pacs, and \keywords
\maketitle

% body of paper here - Use proper section commands
% References should be done using the \cite, \ref, and \label commands
\section{Introduction}
% Put \label in argument of \section for cross-referencing
%\section{\label{}}
%$\subsection{}
%$\subsubsection{}
Filled skutterudites with a chemical formulae RT$_4$X$_{12}$, where R is a rare earth, T transition metal and X=P, As, or Sb, have been lately of great interest due to their broad spectrum of exotic properties related to the unique crystal structure.  Among them, PrOs$_4$Sb$_{12}$ is the first heavy fermion superconductor based on Pr.\cite{Bauer} PrFe$_4$P$_{12}$ is a field induced heavy fermion compound.\cite{Aoki} CeRu$_4$Sb$_{12}$ is a non-Fermi liquid system.\cite{Takeda} CeOs$_4$Sb$_{12}$ is a possible Kondo insulator.\cite{Bauer2} This last material is the least studied and the interpretation of its properties is controversial.

The resistivity of CeOs$_4$Sb$_{12}$ is metallic at room temperature, but shows behavior typical of an insulator, i.e., increases strongly with a decrease of temperature, below 50 K.  The resistivity does not follow an activation-type temperature variation. However, such a variation would be expected only at temperatures much lower than the energy gap of approximately 10 K.\cite{Bauer2} There is an evidence that under the hydrostatic pressure of order 8 GPa, the resistivity evolves into a variable-range hoping type.\cite{Hedo} A recent study of optical conductivity spectra\cite{Matsunami} shows a strong temperature dependence and a gap of 30 meV below 60 K. In addition, a structure observed at 70 meV was interpreted in terms of the hybridization gap peak. These transport properties are quite consistent with a Kondo insulator or hybridizataion gap semiconductor.\cite{Aeppli}

On the other hand, the two reported values of the electronic specific heat coefficient, $\gamma$, are 90\cite{Bauer2} and 180 mJ/K$^2$mol\cite{Namiki}, hardly consistent with the presence of an energy gap. The specific heat also exhibits an anomaly near 1.1 K. Because of a very small entropy associated with this anomaly (about 2 \% of Rln2 only), it has been initially ascribed to some impurity phase\cite{Bauer2}. However, the subsequent magnetic field study\cite{Namiki} up to 4 T has undermined this original interpretation.

To shed more light on the nature of the ground state in CeOs$_4$Sb$_12$, we have extended the specific heat measurements to fields as large as 14 T and introduced La-impurities on Ce sites.

\section{Experimental and Results}

Single crystalline samples of Ce$_{1-x}$La$_x$Os$_4$Sb$_{12}$, with $x=0, 0.02,$ and 0.1, were obtained by the Sb-self flux method described elsewhere.\cite{Bauer2} For mixed alloys, Ce and La were premelted first to assure the homogeneity of samples. X-ray diffraction measurements were performed on powdered single crystals using Philips APD 3720 diffractometer. All spectra were indexed by the BCC (${\it Im}\overline{3}$) crystal structure. No external phases were detected. The change of the lattice constants between the end-compounds, CeOs$_4$Sb$_{12}$ and LaOs$_4$Sb$_{12}$, is very small is very small and on the edge of our experimental resolution. These lattice constants are, 9.304(2) and 9.309(2) {\AA}, respectively. On the other hand, any variation of the lattice constant between $x=0$ and 0.1 is too small to be reliably resolved by this technique. This insensitivity of the lattice constant of ROs$_4$Sb$_{12}$ to the atomic size of R is a unique property of this crystal structure. \cite{Jeitschko,Bauer2} A Ce ion in CeOs$_4$Sb$_{12}$ is surrounded by an oversized icosahedral cage formed by 12 Sb ions. Thus, only small variations of the hybridization parameters upon La-doping are expected.   

The results of the magnetic field study of the pure compound are shown in Fig. 1.  The comparison of all up-to-date specific heat measurements indicates strong sample dependency. Our data, in the form of $C/T$ (specific heat divided by temperature) versus $T$, exhibit a peak at $T_m$=1.1 K, thus at a temperature identical to that reported by Bauer et al\cite{Bauer2}. Namiki et al.\cite{Namiki}, on the other hand, found this peak at 0.9 K. Furthermore, the height of the peak is approximately 220, 510,  and 380 mJ/mol K$^2$ according to Bauer et al., Namiki et al., and our data, respectively. There is also a large discrepancy in $C/T$ results above the anomaly. Our data above 3 K are consistent with those of Namiki et al., yielding an electronic specific heat coefficient of 180 mJ/mol K$^2$. This high temperature $\gamma$ is reduced by the 1.1 K transition to a value smaller than 100 mJ/mol K$^2$. A linear extrapolation of our lowest temperature data (in the form of $C/T$ versus $T$) implies that this low temperature $\gamma$ might be in fact 0. However, lower temperature data are needed to make a more quantitative prediction.  Note that results presented in Fig. 1 exclude a nuclear contribution. This nuclear contribution can be separated easily from the electronic one in the relaxation-method calorimetry, if the two characteristic time constants entering the measurement, $\tau$ characterizing the coupling between electronic degrees of freedom and the heat reservoir and the nuclear spin-lattice relaxation time $T_1$, determining the coupling between nuclear and electronic degrees of freedom, differ substantially.\cite{Andraka} Because of a small mass of the single crystal (about 5 mg), this first time constant ($\tau$ $\sim$ 0.1 s) was more than the order of magnitude smaller than $T_1$.

This strong impact of the transition on the linear specific heat coefficient is an obvious indication that either the transition itself is intrinsic to CeOs$_4$Sb$_{12}$ or that both the anomaly and large high temperature $\gamma$ are due to an extrinsic phase. This second scenario is unlikely since this extrinsic phase would have to be a much larger fraction of the sample than initially assumed.\cite{Bauer2} A similar conclusion was reached by Namiki et al., using the specific heat data at 4 T and a similar argument of the reduction of $\gamma$ at $T_m$. An additional evidence supporting the first scenario is provided by the impurity study discussed below. Thus the 1.1 K transition leads to a further (possibly full) opening of the gap at the Fermi energy.

\begin{figure}
\includegraphics{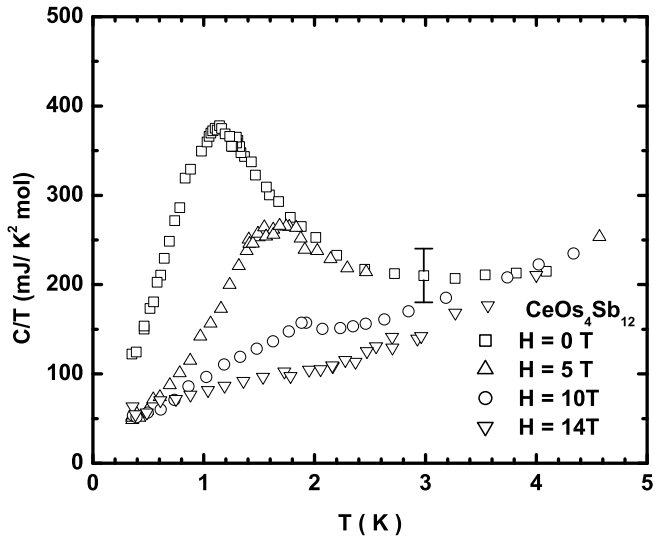}
\caption{$C/T$ versus $T$ for CeOs$_4$Sb$_{12}$ in H=0, 5, 10, and 14 T. Magnetic field was applied along (100) direction. \label{1}}
\end{figure}

In agreement with previous reports,\cite{Namiki,Hitoshi} we find that a moderately strong magnetic field (5 T; applied along the (100) crystallographic direction) shifts the anomaly to a higher temperature, but at the same time, the magnitude of the peak is reduced. Again, the extrapolation of the lowest temperature data to $T=0$ yields a small electronic specific heat coefficient, much smaller than that corresponding to the paramagnetic state. This field has a negligible effect on the high temperature $\gamma$. The anomaly at 10 T is very small, but there is a distinct change of the slope of $C/T$ versus $T$ data at $T_m$, implying a reduction of $\gamma$. On the other hand, we do not detect, within the resolution of our measurement, any anomaly at 14 T. 
 
The field dependence of the specific heat is puzzling for the Kondo insulator scenario.\cite{Aeppli} Considering the data above the anomaly only, we find a sizable decrease of the Sommerfeld coefficient with a field larger than 5 T. In a typical Kondo insulator the opposite is observed.\cite{Takabatake} Magnetic field destroys the hybridization between f- and conduction-electrons that should result in the closing of the gap and an increase of $\gamma$. On the other hand, our $C/T$ data below $T_m$, extrapolated to $T=0$, suggest that there is a slight increase of the low temperature $\gamma$ with magnetic field. This trend is clear upon a closer inspection of the data below 1 K for fields 5, 10, and 14 T.

\begin{figure}
\includegraphics{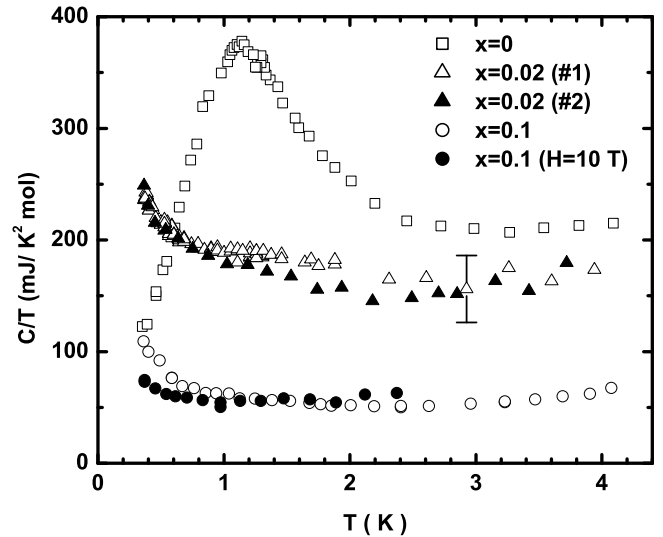}
\caption{$C/T$ versus $T$ for Ce$_{1-x}$La$_x$Os$_4$Sb$_{12}$; where $x=0$, 0.02, and 0.1 (H=0 and 10 T). \label{2}}
\end{figure}

This low temperature state of  CeOs$_4$Sb$_{12}$ is extremely sensitive to La-impurities. Just 2 \% of La introduced for Ce suppresses the anomaly to temperatures lower than 0.4 K. (Fig. 2) There is an increase of $C/T$ below 1 K, possibly related to this transition taking place at much lower temperatures. There is also a relatively large decrease of the electronic specific heat in the paramagnetic state. In order to confirm this result we have measured two different crystals with different masses. These samples were approximately 7 mg (\#1) and 2 mg (\#2). Large scattering was due to a combination of a short time constant ($\tau$), large background contribution, and a lack of averaging of several measurements at the same temperature (performed for $x=0$ and 0.02). Nevertheless, there is a good agreement between the two sets of data for $x=0.02$, demonstrating that our addenda subtraction is accurate. Moreover, this addenda is of the same magnitude for the pure sample and sample \#1, thus can not explain a large drop in $C/T$ between $x=0$ and 0.02. Further substitution of La for Ce results in additional reduction of the specific heat. This reduction, say at 3 K, is by more than 70 \% for just 10 \% of La. The 10 T magnetic field applied along the (100) direction suppresses the low temperature tail and reduces further $\gamma$ obtained from the extrapolation of $C/T$ versus $T$ data to $T=0$ (See Fig. 2 for 10 T data.)

A disappearance of the 1.1 K peak in the La-doped samples provides a complementary evidence for the intrinsic origin of this peak. La is chemically similar to Ce but carries no f-electron. The substitution of La for Ce leads to  some randomness on f-electron sites. The low temperature structure seems to be characteristic of undoped, high quality samples only.

A similarly critical dependence on chemical disorder on the f-ion site was reported for CeAl$_3$.\cite{Andraka2} In this latter case, however, La induces bulk magnetism, as opposed to the case of CeOs$_4$Sb$_{12}$.

\begin{figure}
\includegraphics{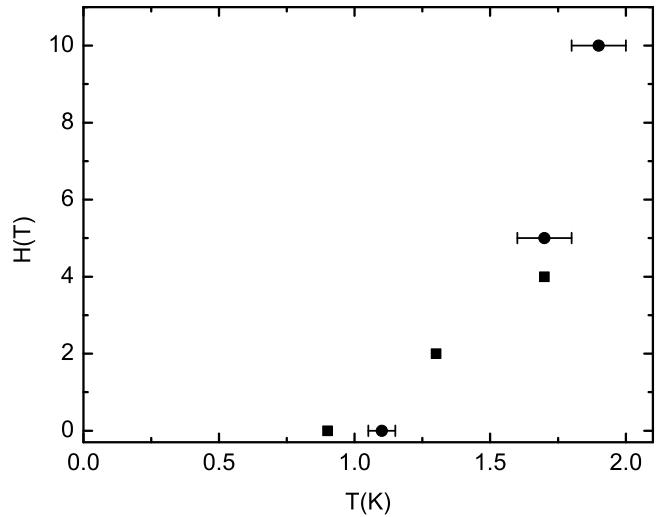}
\caption{Magnetic field phase diagram of CeOs$_4$Sb$_{12}$. The transition temperatures were identified by the maximum values of C/T in Fig. 1 (dots). Squares represent the results of Namiki et al.\cite{Namiki} \label{3}}
\end{figure}

\section{Conclusions}

Our results point to a number of unusual features of the low temperature behavior of CeOs$_4$Sb$_{12}$.  First of all,  the magnetic phase diagram shown in Fig. 3 is atypical of a Ce-based Kondo insulator or a heavy fermion system. A positive sign of $dT_m/dH$ is inconsistent with antiferromagnetism. The temperature variation of the susceptibility indicates the predominance of aniferromagnetic-type interactions rather than ferromagnetic.\cite{Bauer2} Also, a low value of the susceptibility at 1.8 K argues against ferromagnetic character of the ordered state.  As it was noticed by Namiki et al., this phase diagram is reminiscent of CeB$_6$\cite{Fujita} (or La-doped\cite{Hiroi} CeB$_6$), which undergoes antiferroquadrupolar order. However, this type of order is inconsistent with a small entropy released below $T_m$ and with the proposed crystalline electric field (CEF) configuration of Ce in CeOs$_4$Sb$_{12}$. The CEF ground state of Ce$^{3+}$ in a cubic environment can be either $\Gamma_7$ doublet or $\Gamma_8$ quartet. A fit of the susceptibility versus temperature seems to indicate that $\Gamma_7$ is the lowest energy state.\cite{Bauer2} Furthermore, the excited $\Gamma_8$ is 327 K above the CEF ground state. However, $\Gamma_7$ does not possess a quadrupolar electric moment. Thus, neutron scattering measurements are needed to verify the CEF configuration of Ce in this compound.

Our results demonstrate a strong correlation between the electronic specific heat coefficient of the paramagnetic state and the presence of the low temperature anomaly. Less perfect samples, containing small amounts of La, have both reduced $\gamma$ and suppressed transition to the ordered state with respect to the pure compound. This behavior strongly suggests that enhanced $\gamma$ is an intrinsic property of CeOs$_4$Sb$_{12}$ and that the 1.1 K transition has itinerant character. Small entropy removed is also consistent with this itinerant character of the transition. Therefore, the reported insulating-like resistivity remains a puzzle.  
\begin{acknowledgments}
This work has been supported by the U.S. Department of Energy, Grant No.
DE-FG02- 99ER45748.
\end{acknowledgments}

% Create the reference section using BibTeX:
\bibliography{CeOs}
\end{document}